\documentclass[manuscript,screen]{acmart}

\usepackage[utf8]{inputenc} %
\usepackage[T1]{fontenc}    %
\usepackage{hyperref}       %
\usepackage{url}            %
\usepackage{booktabs}       %
\usepackage{amsfonts}       %
\usepackage{nicefrac}       %
\usepackage{microtype}      %
\usepackage{xcolor}         %
\usepackage{amsmath}
\usepackage{natbib}

\usepackage{subfigure}
\usepackage{algorithmic}
\usepackage{multirow}
\usepackage{graphicx}
\usepackage{textcomp}
\usepackage{enumitem}
\usepackage{comment}
\usepackage{pifont}%
\newcommand{\cmark}{\ding{51}}%
\newcommand{\xmark}{\ding{55}}%

\usepackage{setspace}
\usepackage{wrapfig,lipsum}

\newcommand{\hl}[1]{\textcolor{black}{#1}}

\setcopyright{acmlicensed}
\copyrightyear{2025}
\acmYear{2025}
\acmDOI{XXXXXXX.XXXXXXX}

\acmConference[TOIS]{Transactions on Information Systems}{June 03--05,
  2018}{Woodstock, NY}
\acmISBN{978-1-4503-XXXX-X/18/06}

\begin{document}
\title{Disentangled Interest Network for Out-of-Distribution CTR Prediction}

\author{Yu Zheng}
\affiliation{%
  \institution{Tsinghua University}
  \city{Beijing}
  \country{China}}
\email{zhengyu.davy@foxmail.com}

\author{Chen Gao}
\authornote{Corresponding authors.}
\affiliation{%
  \institution{Tsinghua University}
  \city{Beijing}
  \country{China}}
\email{chgao96@tsinghua.edu.cn}

\author{Jianxin Chang}
\affiliation{%
  \institution{Kuaishou}
  \city{Beijing}
  \country{China}}
\email{jianxin.chang@outlook.com}

\author{Yanan Niu}
\affiliation{%
  \institution{Kuaishou}
  \city{Beijing}
  \country{China}}
\email{alanniu@outlook.com}

\author{Yang Song}
\affiliation{%
  \institution{Kuaishou}
  \city{Beijing}
  \country{China}}
\email{yangsong@kuaishou.com}

\author{Depeng Jin}
\affiliation{%
  \institution{Tsinghua University}
  \city{Beijing}
  \country{China}}
\email{jindp@tsinghua.edu.cn}

\author{Meng Wang}
\affiliation{%
  \institution{Hefei University of Technology}
  \city{Hefei}
  \country{China}}
\email{eric.mengwang@gmail.com}

\author{Yong Li}
\authornotemark[1]
\affiliation{%
  \institution{Tsinghua University}
  \city{Beijing}
  \country{China}}
\email{liyong07@tsinghua.edu.cn}

\begin{abstract}
    Click-through rate (CTR) prediction, which estimates the probability of a user clicking on a given item, is a critical task for online information services. Existing approaches often make strong assumptions that training and test data come from the same distribution. However, the data distribution varies since user interests are constantly evolving, resulting in the out-of-distribution (OOD) issue. In addition, users tend to have multiple interests, some of which evolve faster than others. 
    \hl{Towards this end, we propose Disentangled Click-Through Rate prediction (DiseCTR), which introduces a causal perspective of recommendation and disentangles multiple aspects of user interests to alleviate the OOD issue in recommendation. We conduct a causal factorization of CTR prediction involving user interest, exposure model, and click model, based on which we develop a deep learning implementation for these three causal mechanisms.} Specifically, we first design an interest encoder with sparse attention which maps raw features to user interests, and then introduce a weakly supervised interest disentangler to learn independent interest embeddings, which are further integrated by an attentive interest aggregator for prediction. \hl{Experimental results on three real-world datasets show that DiseCTR achieves the best accuracy and robustness in OOD recommendation against state-of-the-art approaches, significantly improving AUC and GAUC by over 0.02 and reducing logloss by over 13.7\%}. Further analyses demonstrate that DiseCTR successfully disentangles user interests, which is the key to OOD generalization for CTR prediction. We have released the code and data at https://github.com/DavyMorgan/DiseCTR/.
    
\end{abstract}



\maketitle

\section{Introduction}\label{sec::intro}
\begin{figure}[t]
    \centering
    \includegraphics[width=\linewidth]{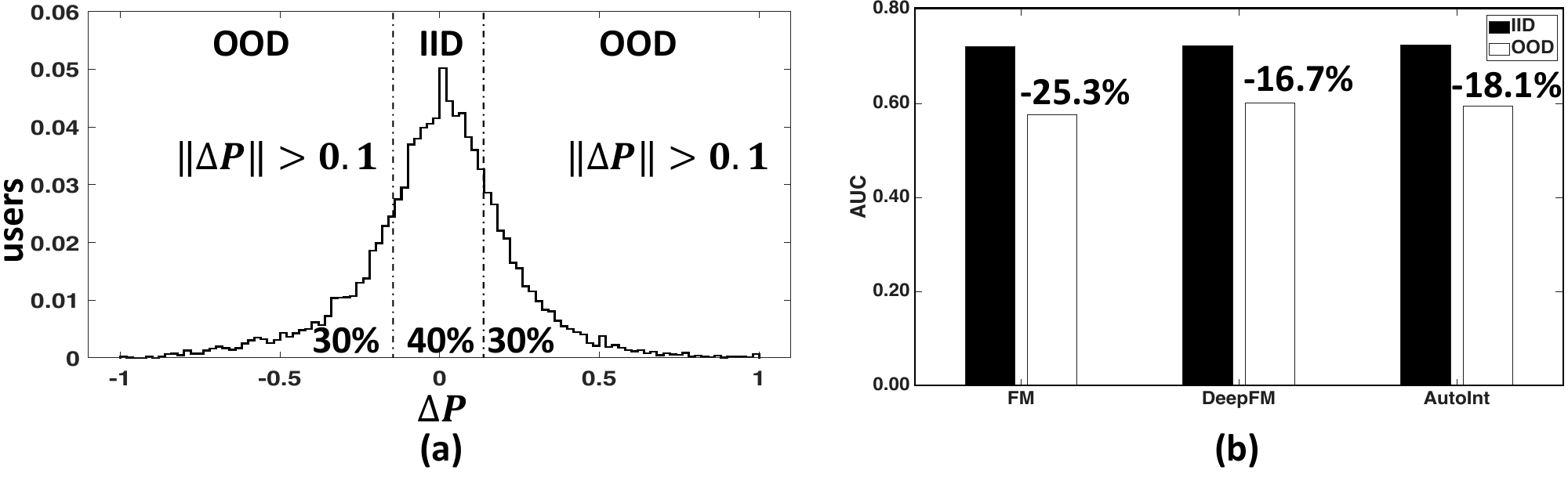}
    \caption{(a) OOD issue on a video recommendation dataset shown with histogram of $\Delta P$, where $\Delta P = P_{eval}(Y=1|X) - P_{train}(Y=1|X)$ is the difference of CTR on popular videos between training and evaluation data. About 60\% of users have a distribution variation larger than 0.1, and only 40\% of the data can be considered as IID. (b) Performance of three popular CTR prediction models on both IID and OOD data, showing their degraded performance when OOD issue occurs.}
    \label{fig::motivation}
\end{figure}

Click-through rate (CTR) prediction plays a crucial role in today's online information services, including recommender systems \citep{cheng2016wide,qin2023learning,gao2023cirs,qin2023learning2,meng2023coarse,yan2023cascading,wang2023sequential,chen2020efficient}, information retrieval \citep{zhang2023user,liu2019search,li2021seamlessly} and computational advertising \citep{song2019autoint}, which takes attributes of users and items as input, and predicts the probability of user interacting with a given item.
In general, CTR prediction can be regarded as a classification task on a binary target variable $Y$ (interaction) from a high-dimensional variable $X$ (user ID, item ID, and features), which has been widely studied in both academia and industry \citep{rendle2010factorization,cheng2016wide,he2017neural,song2019autoint,cheng2020adaptive,qin2020user,pi2019practice,zhu2021disentangled}.
Following classical machine learning theories \citep{mohri2018foundations}, existing CTR models rely on the assumption that training and evaluation data are \textit{independent and identically distributed (IID)}.
Particularly, they assume that the conditional distribution $P(Y|X)$ remains stable.
In practice, however, the data distribution tends to drift over time, causing distribution discrepancies between training and test data, which is known as the \textit{out-of-distribution (OOD)} issue~ \citep{arjovsky2019invariant,hendrycks2021many,ye2021towards,shen2021towards,lu2021nonlinear,lu2022invariant}.
For example, the CTR of a football-related video is often high within the first few days/hours when the match concludes, and then declines sharply over time.
To illustrate the distribution variation, we conduct an analysis on a real-world video recommendation dataset. (details of the dataset will be introduced in Section \ref{sec::exp}).
Specifically, for each user in the dataset, we calculate the differences of CTRs $P(Y|X)$ on {\em popular} videos between training and test data, denoted as $\Delta P$ for short.
The histogram of $\Delta P$ for all users is shown in Figure \ref{fig::motivation}(a).
We can observe that the IID assumption only holds for about 40\% of the users, while the remaining 60\% of the users have large distribution variation with $\|\Delta P\|$ greater than 0.1, which indicates the severe OOD issue.
In other words, in real-world scenarios, the OOD issue is quite common and the IID assumption no longer holds.
As a result, the performance of existing CTR models trained under the IID assumption can no-longer be guaranteed.
To illustrate that, we trained three popular CTR prediction models (FM \citep{rendle2010factorization}, DeepFM \citep{guo2017deepfm} and AutoInt \citep{song2019autoint}), and evaluated their performance with both IID and OOD data.
Figure \ref{fig::motivation}(b) shows that the AUC of all three models drops significantly in OOD scenarios.
Therefore, it is critical to develop a robust recommender system that can generalize well under OOD circumstances.

To approach the problem, we first investigate the sources of distribution variation and why existing CTR models fail.
From the view of users, they are not interacting with items based on \textit{low-level features}.
In real-world recommendation, users tend to have multiple unobserved \textit{interests} ($Z=\{Z_1, \cdots, Z_M\}$), such as aesthetic-aspect interest, social-aspect interest, and economic-aspect interest, which determines whether a user will interact with an item ($Y$).
Each interest can be related to a few raw features ($X$), \textit{e.g.}, the economic-aspect interest is largely related to the price feature and brand feature.
When the users browse the items, the interests rather than features, are the intrinsic drives of user-item interactions.
Formally, there exists a set of unobserved interest variables $Z=\{Z_1, \cdots, Z_M\}$, each of which is related to a certain part of features, determining whether a user will interact with an item.
Meanwhile, interests usually change in a \textit{partial} manner, \textit{i.e.}, only a few vary and the majority remains stable \citep{scholkopf2019causality,scholkopf2021toward}.
There is a commonly existing example: \textit{giving coupons to e-commerce users tends to change their economic interest, while other interests are largely stable} \citep{guo2019buying,li2020spending}.
In other words, the partial changes on $Z$ indicate small variation of $P(Z|X)$ and $P(Y|Z)$, while the resulted shift of $P(Y|X)$ can be large, since one affected $Z_i$ can be related to many features.
Here we follow the above example, and there will be: \textit{change of economic interest can influence features like price, brand, and category.}
As a result, existing CTR models fail to generalize in OOD scenarios since they ignore $Z$ and directly capture $P(Y|X)$ which is vulnerable to the partial distribution variation on $Z$.

\hl{To address the OOD issue, a CTR model can take advantage of the partial property by learning disentangled interests.
Specifically, based on a causal factorization of recommendation decomposing the joint distribution of features, interests, and interactions into an interest model $P(Z)$, an exposure model $p(X|Z)$, and a click model $P(Y|X,Z)$, we can capture $P(Z|X)$ and $P(Y|Z)$ separately rather than directly learning $P(Y|X)$.
In this way, the impact of distribution variation is reduced to a few affected interests, and most parts of the model will not be influenced.
Thus it can generalize to OOD scenarios efficiently with only the affected part being updated.
However, learning disentangled interests is non-trivial with the following challenges.
First, interests are unobserved causal variables that implicitly correspond with input features.
There is no label for each $Z_i$ as well as which features are related to $Z_i$.
Second, distribution variation can independently occur on any $Z_i$, thus it is necessary to decouple different interests to ensure that the model can handle all possible changes.}

\begin{figure}
    \centering
    \includegraphics[width=\linewidth]{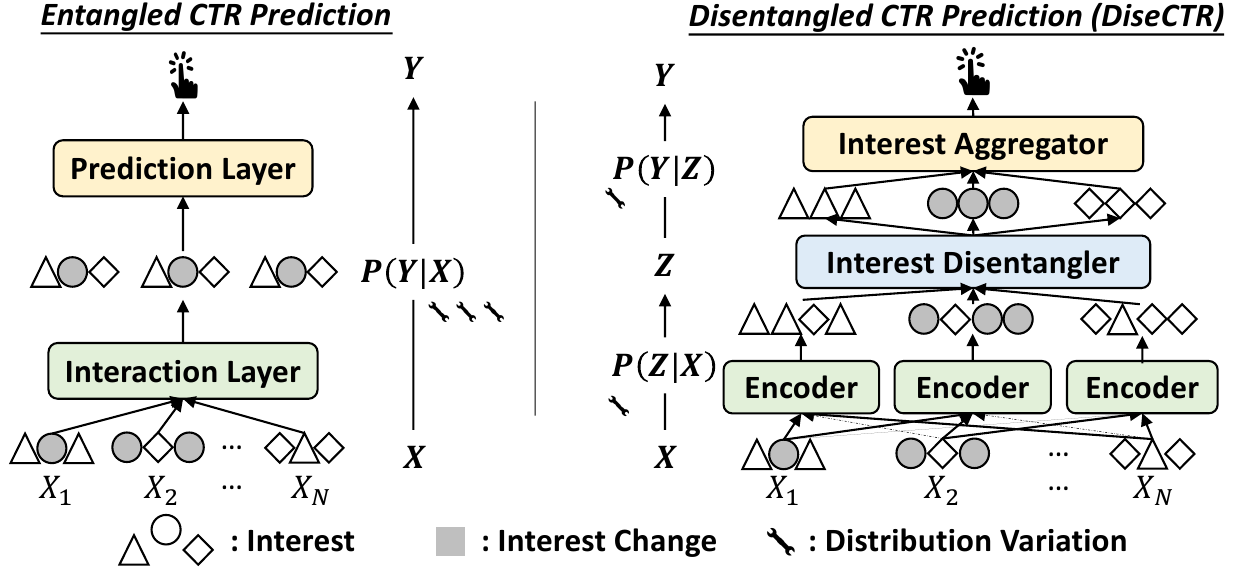}
    \caption{Existing entangled methods directly learn $P(Y|X)$, while DiseCTR learns $P(Z|X)$ and $P(Y|Z)$ separately with disentangled interests. Under partial changes of interests (gray in the figure), existing methods tend to fail due to the large variation of $P(Y|X)$, while DiseCTR is more robust since variation of $P(Z|X)$ and $P(Y|Z)$ is relatively small.}
    \label{fig::method}
\end{figure}

It is worth noting that two recent works \cite{he2022causpref,wang2022causal} investigated the OOD issue in recommendation, however, they focused on simplified scenarios (collaborative filtering \cite{he2022causpref} and user income increase \cite{wang2022causal}), and they can not handle general OOD issues in CTR prediction.
In this paper, we propose a disentangled interest network named DiseCTR to achieve OOD CTR prediction under both simple and complicated distribution variation.
To address the above-mentioned challenges, we first design an interest encoder that utilizes sparse attention \citep{zhou2021informer} to achieve effective interest embeddings, of which each interest only involves a few related features.
We then propose a weakly supervised interest disentangler which can regularize each interest to be both meaningful and independent.
With the encoder and disentangler, we can generate robust interest embeddings from low-level features, which can well capture $P(Z|X)$.
We then propose an attentive interest aggregator to accomplish $P(Y|Z)$, combining different interests for the final prediction.
Overall speaking, DiseCTR takes full advantage of the \textit{partial-distribution-variation} property, and thus it can generalize better towards OOD scenarios compared with existing approaches of CTR prediction.
We illustrate the critical differences between DiseCTR and existing methods in Figure \ref{fig::method}.
To evaluate the accuracy and robustness of DiseCTR, we experiment on three real-world datasets.
\hl{Results show that DiseCTR outperforms state-of-the-art approaches with significant improvements by over 0.02 on AUC and GAUC (improvement of 0.001 is acknowledged as significant \citep{song2019autoint,cheng2020adaptive}) and over 13.7\% on logloss}.
Moreover, we demonstrate the superior robustness of DiseCTR in OOD scenarios, especially for large distribution variation.
Further in-depth analyses empirically show that DiseCTR successfully disentangles user interests, which is crucial to OOD generalization.

The contribution of this paper can be summarized as follows.
\begin{itemize}
    \item We highlight the OOD issue and the partial-distribution-variation property in recommendation, and take the pioneer step to study OOD generalization for the general CTR prediction task.
  \item \hl{We design a novel model DiseCTR which introduces a causal perspective to disentangle user interests for OOD CTR prediction. The proposed method captures the partial-variation property with a sparse attentive interest encoder and a weakly supervised interest disentangler.}
  \item We conduct extensive experiments on three real-world datasets.
  Experimental results demonstrate the superior performance of DiseCTR against state-of-the-art CTR models. 
\end{itemize}

The remainder of the paper is as follows.
We first formulate the problem of robust CTR prediction in Section \ref{sec::probdef} and elaborate on our proposed DiseCTR in Section \ref{sec::method}. 
We then conduct experiments in Section \ref{sec::exp} and carefully review related works in Section \ref{sec::related}. 
\hl{Last, we conclude the paper and discuss the limitations and future works in Section \ref{sec::conclusion}.}

\section{Problem Formulation}\label{sec::probdef}

The CTR prediction problem is formulated as estimating a binary target $Y$ from a feature vector $\mathbf{X}$ that is composed of $N$ fields $\{X_1, X_2, \cdots, X_N\}$.
Let $\mathcal{D}$ denote the joint distribution, $P(X,Y)$.
We aim at learning a CTR model on training data $\mathcal{O}_{train}$, that can generalize to the test data $\mathcal{O}_{test}$ with $\mathcal{D}_{test}$ different from $\mathcal{D}_{train}$.
Then the problem can be formulated as follows, \\
\textbf{Input:} Training dataset $\mathcal{O}_{train}$ and OOD test dataset $\mathcal{O}_{test}$, which both contain users, items and their attributes as feature vectors $\mathbf{X}$, and corresponding interaction labels $Y$.\\
\textbf{Output:} A model estimating the label with strong generalization ability towards unseen distributions.

It is worthwhile noting that the generalization ability of a CTR model is measured comprehensively following existing literature \cite{bengio2019meta}, which includes both accuracy and efficiency of transferring to OOD senarios, \textit{i.e.} $\mathcal{O}_{test}$.
as follows,
\begin{itemize}[leftmargin=*]
    \item \textbf{Transfer Accuracy.} The accuracy after transferring to $\mathcal{O}_{test}$.
    \item \textbf{Transfer Efficiency.} The sample cost of transferring to $\mathcal{O}_{test}$.
\end{itemize}

\section{Method}\label{sec::method}

\hl{\subsection{A causal view of CTR prediction}
It is acknowledged that high-dimensional observations such as images, are the causal effect of a set of low-dimensional independent factors ~\citep{peters2017elements,pearl2018book,scholkopf2019causality,bengio2019meta,scholkopf2021toward}.
For example, an image of a cube containing thousands of pixels can be fully determined by a few causal variables such as the position and rotation of the cube.
Therefore, learning representations that can keep consistent with the underlying causal mechanisms is expected to achieve better OOD generalization \citep{scholkopf2019causality,scholkopf2021toward,gao2024causal}.
It can be explained that distribution variation usually comes from real-world interventions, which tend to influence only a few causal variables.
In this paper, we view CTR prediction from the perspective of causal inference, and develop a causal modeling of interests, features, and interactions in recommendation.
Specifically, building upon well-established causal theories~\citep{peters2017elements,pearl2018book,scholkopf2019causality,bengio2019meta,scholkopf2021toward}, DiseCTR first decodes the underlying causal variables (interests) from high-dimensional observations (features), and then infer the effect (interaction) of these causal variables.}

\begin{figure}[h]
    \centering
    \includegraphics[width=\linewidth]{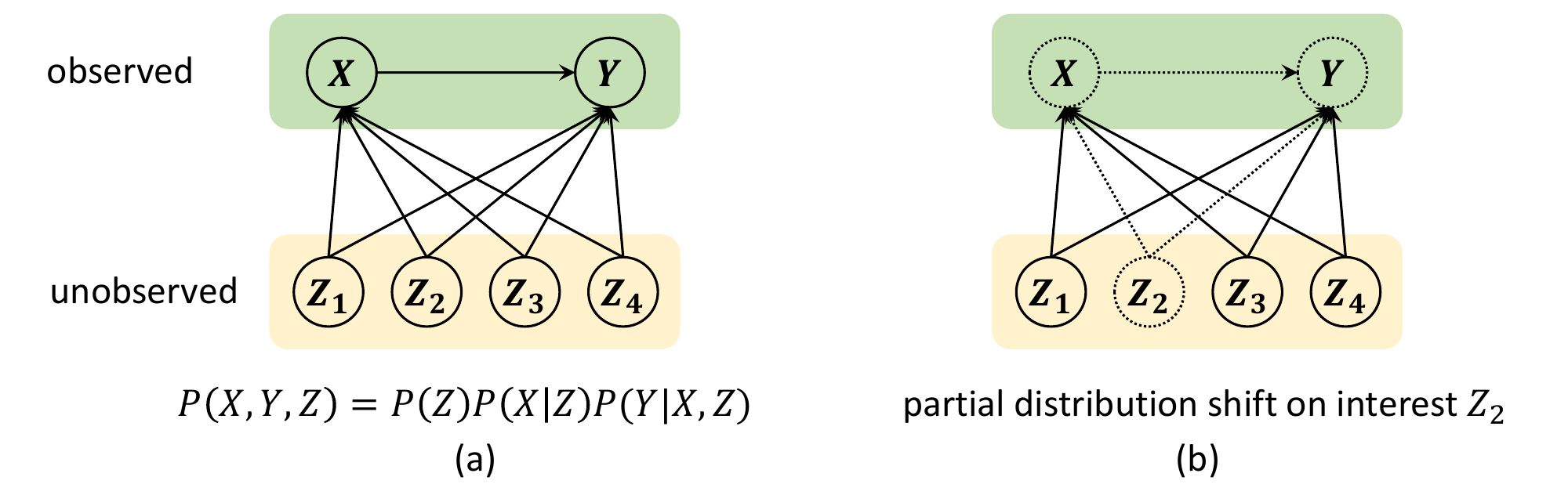}
    \caption{(a) Causal graph of features $X$, interactions $Y$ and interests $Z$. (b) Partial distribution shift .}
    \label{fig::causal}
\end{figure}

\hl{Intuitively, interactions between users and recommender systems can be modeled as a causal process of three types of causal variables, which are interests $Z$, features $X$ and interactions $Y$.
Figure \ref{fig::causal}(a) illustrates the causal graph, where we show four interests for example.
The causal factorization \citep{peters2017elements,scholkopf2019causality} of the joint distribution $P(X, Y, Z)$ is composed of three causal mechanisms, which are the intrinsic distribution of user interests $P(Z)$, the exposure model $P(X|Z)$ and the click model $P(Y|X,Z)$.
Firstly, $P(Z)$ describes the user interests which are unobserved causal variables.
Secondly, what contents are exposed to users is determined by the recommender system which learns to capture user interests.
In other words, what users are interested in ($P(Z)$) decides what users see ($P(X)$) from the recommender system, which is the exposure model $P(X|Z)$ in the causal factorization.
Thirdly, whether a user clicks an item ($P(Y)$) is determined by both what the user see ($P(X)$) and whether the user is interested in the item ($P(Z)$), which is the final click mechanism $P(Y|X,Z)$ in the causal factorization.}

\hl{Most existing CTR prediction approaches directly ignore $Z$ and learn $P(Y|X)$, since user interest $Z$ is unobserved and difficult to track, 
However, such straightforward solutions are vulnerable to distribution shift which is common in real-world applications.
As introduced previously, user interests usually change in a \textit{partial} way, indicating that the variation of $Z$ is sparse.
Figure \ref{fig::causal}(b) demonstrates a sparse user interest shift.
Specifically, in the example, one of the four interests, $Z_2$, changes, while the other three interests ($Z_1$, $Z_3$, and $Z_4$) remain largely stable.
Since $Z_2$ is causally related to both $X$ and $Y$, the conditional distribution $P(Y|X)$ is significantly affected by the drifted interest.
Traditional CTR models directly learn $P(Y|X)$, which is vulnerable to the distribution variation occurring partially on interest variables $Z$.
As a consequence, they tend to suffer large performance drops.}

\hl{Unlike existing CTR predictions, we propose to learn $P(Z|X)$ and $P(Y|Z)$ instead of $P(Y|X)$, grounding DiseCTR on the actual causal factorization of recommendation illustrated by Figure \ref{fig::causal}(a).
Using the above example, since the remaining majority of user interests, $Z_1$, $Z_3$ and $Z_4$, keep largely unchanged, DiseCTR can still take effect under the partial variation on $Z_2$.
By leveraging the partial distribution variation property, DiseCTR is more robust against interventions and can better generalize to OOD scenarios.}

\hl{The proposed DiseCTR consists of the following three main components, interest encoder, interest disentangler and interest aggregator, as shown in Figure \ref{fig::method}.
\begin{itemize}
    \item \textbf{Interest Encoder.} 
    To capture multiple aspects of user interests, we design an encoder to generate a set of interest embeddings.
    We utilize sparse attention~\citep{zhou2021informer} in the encoder which can force each interest to be related to only a small part of input features.
    \item \textbf{Interest Disentangler.} 
    In order to leverage the partial property of interest shift, the obtained embeddings are first clustered to a set of interest prototypes, and further disentangled with pairwise weak supervision.
    In this way, interest embeddings are regularized to be independent and meaningful, which reduces the impact of distribution variation in OOD scenarios.
    \item \textbf{Interest Aggregator.} 
    We design an attentive interest aggregator to predict CTR based on disentangled interests, and the attention weights of different interests are adapted to unseen distributions.
\end{itemize}
From the causal perspective, we first capture $P(Z|X)$ with the interest encoder and enhance robustness using the interest disentangler, then capture $P(Y|Z)$ with the interest aggregator.}

\subsection{Interest Encoder}
Users tend to have multiple high-level interests $Z$ which are implicitly related to low-level features $X$.
Take e-commerce recommendation as an example, $Z_i$ may be content interest related to item category and item image.
While another $Z_j$ may be economic interest correlated with item price and item brand.
In other words, each interest corresponds with only a part of features, and different interests focus on distinct fields.
Inspired by the recent study of sequential modeling that aims at attending to a few dominant timestamps, we utilize a sparse attention based encoder~\citep{zhou2021informer} on input features.
Specifically, we develop multiple sub-encoders, and each one generates an interest embedding by attending to a small fraction of features.
Note that features are transformed into dense vectors by an embedding layer before being fed into sub-encoders.

\noindent\textbf{Embedding Layer.}
Since raw features are usually high dimensional, they are commonly transformed into dense vectors with embedding matrices \citep{guo2017deepfm,song2019autoint} as follows,
\begin{equation}
    \mathbf{e} = [\mathbf{E_1}(x_1); \mathbf{E_2}(x_2); \cdots; \mathbf{E_N}(x_N)],
\end{equation}
where $N$ is the number of features, and $x_i$ is the one-hot encoding of each feature.
Here $\mathbf{E_i} \in \mathbb{R}^{N_i \times d}$ is an embedding matrix, with $N_i$ as the number of possible values of $x_i$ and $d$ as the embedding size.
With the embedding layer, we can express each sample as a tensor of shape $(d, N)$, and each column represents one feature.
Notice that the embedding layer is shared across the following $M$ parallel sub-encoders, thus the number of embedding parameters in DiseCTR is the same with existing CTR models.

\noindent\textbf{Parallel Sparse Sub-encoder.}
We design $M$ parallel sub-encoders to capture multiple interests based on the multi-head self-attention mechanism \citep{vaswani2017attention}, where $M$ is a hyper-parameter.
We first generate query, key, and value for attention as follows,
\begin{equation}
	\textbf{q} = \mathbf{W_qe};~
	\textbf{k} = \mathbf{W_ke};~
	\textbf{v} = \mathbf{W_ve},
\end{equation}
where $\mathbf{W_q}, \mathbf{W_k}, \mathbf{W_v} \in \mathbb{R}^{d' \times d}$ are learnable transformation matrices and $d'=Hd$ with $H$ as the number of attention heads.
In the original full attention \citep{luong2015effective}, we have to compute the similarity between any pair of query and key to obtain attention scores.
However, the learned interest embedding is then related to all the features.
As each interest only corresponds with a small fraction of features, we leverage \textit{ProbSparse} self-attention in Informer \citep{zhou2021informer} 
which only attends to a few dominant features, and the interest embedding is obtained by flattening the output of the attention layer, as follows:
\begin{equation}
    \tilde{\mathcal{A}} = \text{\tt SoftMax}(\frac{\tilde{\mathbf{q}}\mathbf{k}^T}{\sqrt{d}})\mathbf{v},~
    \tilde{\mathbf{z}} = \text{\tt concat}(\{\mathbf{a}_1, \cdots, \mathbf{a}_N\}), \label{eq::concat}
\end{equation}
where $\tilde{\mathbf{q}}$ contains $c$ queries with $c \ll N$, and $a_i$ is the $i$-th row of the attentive output $\tilde{\mathcal{A}}$.
In this way, the interest embedding is forced to be related to only a small part of input features.
Readers can refer to Eqn (2-4) in \citep{zhou2021informer} for details on the computation of Eqn (\ref{eq::concat}).

To capture multiple interests, we develop $M$ sub-encoders to generate a set of interest embeddings,
\begin{equation}
    \tilde{\mathbf{Z}} = [\tilde{\mathbf{z}}_1, \tilde{\mathbf{z}}_2, \cdots, \tilde{\mathbf{z}}_M],
\end{equation}
where each $\tilde{\mathbf{z}}_i \in \mathbb{R}^{d'}$ is generated with Eqn(\ref{eq::concat}). 
Although the sparse attention layer makes each $\tilde{\mathbf{z}}_i$ correspond to only a few features, simply paralleling $M$ sub-encoders is insufficient to capture $M$ different interests, because these sub-encoders can be redundant, and interests in $\tilde{\mathbf{Z}}$ can entangle with each other.
From the perspective of OOD generalization, it is critical to ensure that different interests are decoupled, which can guarantee that distribution variation on one interest will not affect another one.
To this end, we now introduce our proposed disentangler which regularizes different interests in $\tilde{\mathbf{Z}}$ to be both meaningful and independent.

\subsection{Interest Disentangler}

\begin{figure*}[t]
    \centering
    \includegraphics[width=\linewidth]{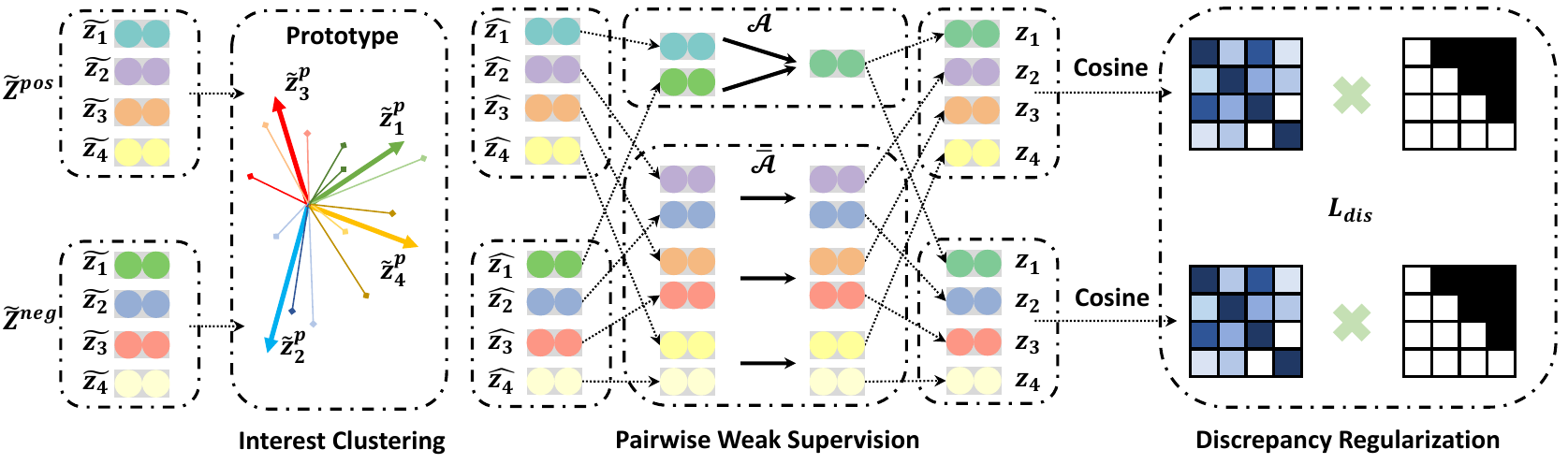}
    \caption{We first cluster $\tilde{\mathbf{Z}}$ towards a set of interest prototypes, then impose weak supervision by averaging the pair adaptively which makes $\mathcal{A}$ and $\bar{\mathcal{A}}$ capture the overlapped and distinguishing interests of the pair respectively. Cosine similarity loss is further minimized to reduce mutual information between different interests. (Best viewed in color).}
    \label{fig::supervision}
\end{figure*}

The output embeddings $\tilde{\mathbf{Z}}$ in Eqn (5) are with high variance since they are directly computed from input features.
Intuitively, interests are several different directions in a latent space, while each sample is a different combination of these directions.
Therefore, we construct a set of interest prototypes which serve as these directions in the latent space, and perform interest clustering to obtain the portions in each direction.
After that we regularize the clustered interests to be both meaningful and independent with pairwise weak supervision and explicit constraints.
Figure \ref{fig::supervision} briefly illustrates the design of the proposed interest disentangler.

\noindent\hl{\textbf{Interest Clustering with Prototypes.}
Since there is no ground-truth for the meaning of each interest, we propose to learn a set of interest prototypes $\mathbf{Z}^p$,
\begin{equation}
    \mathbf{Z}^p = [\tilde{\mathbf{z}}^p_1; \tilde{\mathbf{z}}^p_2; \cdots; \tilde{\mathbf{z}}^p_M],\label{eq::proto}
\end{equation}
where $\tilde{\mathbf{z}}^p_i \in \mathbb{R}^{d'}$ resides in the same latent space and shares the same shape with the encoder outputs $\tilde{\mathbf{Z}}$.
These prototypes are learnable parameters, randomly initialized and gradually converged to centroids of user interests through continuous update using CTR prediction data.
Specifically, we cluster interests in $\tilde{\mathbf{Z}}$ towards $\mathbf{Z}^p$ according to the distance between them.
Then embeddings in $\tilde{\mathbf{Z}}$ are assigned to different clusters by performing projection to prototypes in $\mathbf{Z}^p$ as follows,
\begin{equation}
    p_{j|i} = \frac{\text{\tt Norm}(\tilde{\mathbf{z}}_i) \cdot \text{\tt Norm}(\tilde{\mathbf{z}}^p_j)}{\sum_{l=1}^{M}{\text{\tt Norm}(\tilde{\mathbf{z}}_i) \cdot \text{\tt Norm}(\tilde{\mathbf{z}}^p_l)}},
\end{equation}
where $i, j \in {1, 2, \cdots M}$, and $p_{j|i}$ is the projection weight of the $i$-th encoder output to the $j$-th prototype.
The projection weight of $\tilde{\mathbf{z}}_i$ to different prototypes is normalized to sum to 1.
Notice that we perform L2 normalization ($\text{\tt Norm}(\cdot)$) on encoder outputs and prototypes before calculating the inner product between them.
In other words, we cluster interest embeddings according to the \textit{cosine similarity} with prototypes, which is shown to be effective in preventing mode collapse \citep{ma2019learning,ma2020disentangled}, \textit{i.e.} only a few prototypes are active.
Particularly, using inner product can easily fall into mode collapse, since there is high probability that all the interest embeddings are clustered to the same prototype with the largest norm.
It is worthwhile to notice that the number of prototypes is not necessarily the same with the number of parallel sub-encoders, $M$, while we keep them consistent for simplicity and leave different setups of encoders and prototypes for future work.
}

The clustered interest embeddings are projected from encoder outputs as follows,
\begin{equation}
    \hat{\mathbf{z}}_k = \sum_{i=1}^{M}{p_{k|i} \cdot \tilde{\mathbf{z}}_i}, 
    ~\hat{\mathbf{Z}} = [\hat{\mathbf{z}}_1; \hat{\mathbf{z}}_2; \cdots; \hat{\mathbf{z}}_M],\label{eq::projection}
\end{equation}
where $k = 1, 2, \cdots M$.
By projecting to prototypes, the clustered interest embeddings $\hat{\mathbf{Z}}$ are more stable with lower variance than the original encoder outputs $\tilde{\mathbf{Z}}$, as shown in Figure \ref{fig::supervision}.
Meanwhile, the projection is fully compatible with back-propagation. 
Notice that adding $M$ learnable prototypes only introduce $Md'$ extra parameters, which is very small compared with parameters of embedding matrices.
With such a small amount of parameters, interest embeddings are gathered in several directions specified by prototypes, and we can further regularize them to be meaningful and independent.

\noindent\textbf{Pairwise Weak Supervision.}
In order to capture meaningful interests, we impose explicit supervision on the obtained $M$ embeddings.
Specifically, inspired by recent advances in Variational Auto-Encoder (VAE) \citep{locatello2019disentangling,locatello2020weakly}, we feed the model with samples in a pair-wise manner (a positive and a negative sample), and each pair can share a few interests in common.
Then we regularize a subset of $\hat{\mathbf{Z}}$ to capture the shared interests, and make the rest of $\hat{\mathbf{Z}}$ capture the distinguishing interests of the pair.
For example, a user may click on a cheap mobile phone and skip an expensive one with the same brand and similar functions.
Then this pair is different in terms of economic-aspect interest, while other interests might be similar.
Consequently, we can divide the interests of each pair into two subsets, shared interests $\mathcal{A}$ and distinguishing interests $\bar{\mathcal{A}}$, where $\mathcal{A} \cup \bar{\mathcal{A}} = \{1, 2, \cdots, M\}$ and $\mathcal{A} \cap \bar{\mathcal{A}} = \varnothing$.
Particularly, we can have the following similarity constraints,
\begin{equation}
    \mathrm{sim}(\hat{\textbf{z}}^{pos}_i, \hat{\textbf{z}}^{neg}_i) < \delta_1 < \delta_2 < \mathrm{sim}(\hat{\textbf{z}}^{pos}_j, \hat{\textbf{z}}^{neg}_j), i \in \mathcal{A}, j \in \bar{\mathcal{A}},
\end{equation}
where $pos$ and $neg$ indicate the pair, $\delta_1$ and $\delta_2$ are threshold values, and $\mathrm{sim}(\cdot, \cdot)$ measures cosine similarity.
In other words, it is the interests in $\bar{\mathcal{A}}$ that separate the pair into positive and negative sample, while the interests in $\mathcal{A}$ are not the reasons for the user's different behaviors towards the two items.
In the above example, the economic interest is in $\bar{\mathcal{A}}$, while other interests are in $\mathcal{A}$.

In order to encourage the pair of learned embeddings to share interests in $\mathcal{A}$ and have different interests in $\bar{\mathcal{A}}$, we design the pairwise-similarity based disentangler as follows,
\begin{equation}
    \mathbf{Z}^{pos} = F(\hat{\mathbf{Z}}^{pos}, \hat{\mathbf{Z}}^{neg}, \mathcal{A}),
    \mathbf{Z}^{neg} = F(\hat{\mathbf{Z}}^{neg}, \hat{\mathbf{Z}}^{pos}, \mathcal{A}),
\end{equation}
where $F$ is the disentangling function of the following form,
\begin{align}
    &F(\mathbf{U}, \mathbf{V}, \mathcal{A}) = [f(\mathbf{u}_1, \mathbf{v}_1, \mathcal{A}); \cdots; f(\mathbf{u}_M, \mathbf{v}_M, \mathcal{A})], \\
    &f(\mathbf{u}_i, \mathbf{v}_i, \mathcal{A}) = 
    \begin{cases} 
        (\mathbf{u}_i+\mathbf{v}_i)/2, & \mbox{if } i \in \mathcal{A}; \\
        \mathbf{u}_i,  & \mbox{if } i \in \bar{\mathcal{A}}.
    \end{cases}
\end{align}
Specifically, as illustrated in Figure \ref{fig::supervision}, we replace the interests in $\mathcal{A}$ with the average of the pair, and keep the interests in $\bar{\mathcal{A}}$ unchanged.
As a result, $\mathbf{Z}^{pos}$ and $\mathbf{Z}^{neg}$ now share a few rows in common.
In this way, we assign specific meanings to the learned interests, making embeddings in $\mathcal{A}$ and $\bar{\mathcal{A}}$ capture the shared and distinguishing interests of the sample pair, respectively. 

It is challenging to find the shared interests of a sample pair, thus we propose to determine $\mathcal{A}$ and $\bar{\mathcal{A}}$ adaptively according to the similarity of $M$ interests, which are calculated as follows,
\begin{equation}
    \mathcal{A} = \{i~|~\mathrm{sim}(\hat{\textbf{z}}^{pos}_i, \hat{\textbf{z}}^{neg}_i) < \delta\}, 
    ~\bar{\mathcal{A}} = \{1, 2, \cdots, M\}\setminus\mathcal{A},
\end{equation}
where $\delta$ is set dynamically to make $|\mathcal{A}|=1$ and $|\bar{\mathcal{A}}|=K-1$, \textit{e.g.} $\delta$ can be the second smallest similarity value of the $M$ interests.
In other words, we make the sample pair share one interest, which is rational since it is uncommon to obtain two samples that are different everywhere.
The computation cost of this adaptive approximation is $\mathcal{O}(M)$ which is in \textit{constant time} complexity, and the results turn out to be fine.
From our experiments, the difference between the running time of DiseCTR with and without this adaptive design is less than 0.01\%, thus this adaptive design is very scalable. 
We leave other possible strategies for finding shared interests of sample pairs as future work.

\noindent\textbf{Discrepancy Regularization.}
As introduced previously, distribution variation occurs in a partial way \citep{scholkopf2019causality,scholkopf2021toward}, \textit{i.e.} only a few interests are likely to be intervened while the remaining majority tends to remain stable.
In order to utilize this property, we add a discrepancy constraint which regularizes the similarity between different interests to be as small as possible, making interests independent with each other.
Specifically, we add an loss term on the cosine similarity of each two interests, encouraging $M$ interests to capture different information, formulated as follows,
\begin{equation}
    L_{dis} = \sum_{s \in \{pos, neg\}}{\sum_{i=1}^{M}{\sum_{j=i+1}^M{\text{\tt Norm}(\hat{\mathbf{z}}_i^{s}) \cdot \text{\tt Norm}(\hat{\mathbf{z}}_j^{s})}}}, \label{eq::dis}
\end{equation}
which can be jointly optimized with the main CTR prediction loss.

\noindent\textbf{Remark.}
In short, the interest disentangler encourages embeddings in $\mathbf{Z}$ to be both meaningful and independent.
In terms of OOD generalization, independent embeddings guarantee that only a small part of the model parameter will be influenced by distribution variation, and meaningful embeddings ensure that the remaining parts of the model parameter will still take effect.
Furthermore, such disentanglement can help DiseCTR accomplish fast transfer to OOD scenarios, since only a small affected part needs to be largely updated.
Existing disentangled recommendation approaches \citep{ma2019learning,wang2020disentangled,wang2021learning} only impose independence regularization like the above $L_{dis}$, while they fail to add explicit supervision on the meaning of the disentangled factors.
As a consequence, the learned disentangled representations may capture independent noise signals.
In DiseCTR, we address this problem by introducing weak supervision, which can generate interest representations that are both meaningful and independent.
We will show in experiments that the pairwise weak supervision design is a very crucial component in DiseCTR.

\subsection{Interest Aggregator}
As the interest encoder and disentangler accomplish $P(Z|X)$, we now propose an interest aggregator to achieve $P(Y|Z)$ that combines disentangled interests for prediction.
The key challenge is that interests contribute differently in the original and OOD scenarios.
Specifically, some interests might change in OOD scenarios, thus they become misleading and useless while the importance of other interests increases.
For example, the aesthetic-aspect interest is important when users browse videos to kill time, while it is less important than the social-aspect interest when users browse videos uploaded by friends.
In other words, the aggregation weights of different interests can vary.
Therefore, we design an interest aggregator based on the attention mechanism \citep{luong2015effective} as follows,
\begin{align}
    &\mathbf{A}_{agg} = \text{\tt SoftMax}(\mathbf{W}_{agg}\mathbf{Z}), \\
    &\mathbf{z}_{agg} = \text{\tt Flatten}(\mathbf{A}_{agg}\mathbf{Z}), \\
    &\hat{Y} = \text{\tt sigmoid}(\mathbf{W}_{ctr}\mathbf{z}_{agg}),
\end{align}
where $\mathbf{W}_{agg} \in \mathbb{R}^{h \times d'}$ is a learnable attentive matrix with $h$ as the number of attention heads.
The final output is estimated with a linear predictor $\mathbf{W}_{ctr} \in \mathbb{R}^{hd' \times 1}$.
It is worthwhile noting that the aggregator only introduces $h \times d'$ auxiliary parameters, which makes it easy to transfer to OOD scenarios, especially in practical scenarios where only a few OOD samples are available.
When fine-tuning with OOD samples, $\mathbf{W}_{agg}$ can be updated to re-combine different interests with high efficiency due to its low parameter amount.
Finally, DiseCTR is jointly trained on two objectives, CTR prediction and discrepancy of interests, formulated as follows,
\begin{equation}
    L = L_{bpr}(\hat{Y}^{pos}, \hat{Y}^{neg}) + \alpha L_{dis} + \lambda\|\Theta\|_2,\label{eq::loss}
\end{equation}
where $L_{bpr}$ is the widely adopted Bayesian Personalized Ranking loss \citep{rendle2012bpr}. $\Theta$ denotes the parameters of DiseCTR, and $\alpha, \lambda$ are loss weights for discrepancy and L2 regularization.
The parameter scale of DiseCTR is comparable with existing approaches like AutoInt \citep{song2019autoint}.
From the experiments, DiseCTR can be trained to converge within 20 minutes on a single GPU.

In summary, we design an encoder to obtain a set of interest embeddings, and each is forced to be related to only a few features by sparse attention.
We further propose a disentangler to make interest embeddings independent and meaningful.
At last, we design an attentive aggregator which combines all interests for prediction.
We now provide a discussion on the relation between DiseCTR and existing works from the perspective of OOD generalization.

\subsection{Discussion}

\subsubsection{Comparison between DiseCTR and typical CTR prediction approaches}
The impact of distribution variation is largely reduced by utilizing the partial-variation property, thus DiseCTR can generalize well towards unseen distributions.
From the perspective of model structure, the inner product in Factorization Machine\citep{rendle2010factorization} can be regarded as learning an interest for each feature pair, and the self-attention in AutoInt \citep{song2019autoint} can be considered as learning $N$ interests that each automatically corresponds with $N$ features.
However, these approaches impose no supervision on the learned interests which make them entangle with each other.
In DiseCTR, we encourage different interests to be both meaningful and independent with weak supervision.
Therefore, stronger disentanglement is achieved, which will be further demonstrated in experimental results of Section \ref{sec::exp}.
We summarize the differences between DiseCTR and several typical CTR models in Table \ref{tab::baseline}.

\subsubsection{Relation Between DiseCTR and VAE}
The idea of learning disentangled representations is mostly investigated under the framework of VAE \citep{kingma2013auto,higgins2016beta}, where images are encoded into low-dimensional vectors with independence regularization by minimizing the KL divergence loss.
However, regularizing the independence in such an unsupervised way has been proved to be insufficient to capture meaningful semantics \citep{locatello2019challenging}.
Variants of VAE were proposed which disentangle with weak supervision, either by adding an extra prediction task \citep{locatello2019disentangling}, or by regularizing the similarity between pairs of observations \citep{locatello2020weakly}.
In DiseCTR, discrepancy regularization and pairwise weak supervision have similar effects with these works.
However, DiseCTR extends these methods to a higher dimensional space, \textit{i.e.} DiseCTR learns disentangled vectors while these methods disentangle coordinates in one vector.

\begin{table}
    \caption{Comparison with typical CTR models. We compare the number of features, the number of interests, supervision (shortened as Sup.) on interests and disentanglement (shortened as Disen.) of interests.}
    \label{tab::baseline}
    \begin{tabular}{c|cccc}
      \toprule
      Method & \#Features & \#Interests & Sup. & Disen.  \\
      \midrule
      LR & $N$ & 0 & \xmark & no \\
      FM & $N$ & $\frac{N(N-1)}{2}$ & \xmark & low \\
      AutoInt & $N$ & $N$ & \xmark & low\\
      DiseCTR & $N$ & $M \ll N$ & \cmark & high \\
      \bottomrule
    \end{tabular}
\end{table}

\section{Experiments}\label{sec::exp}

In this section, we conduct extensive experiments to answer the following research questions:
\begin{itemize}
    \item \textbf{RQ1}: What is the overall generalization ability of DiseCTR compared with state-of-the-art CTR prediction approaches?
    
    \item \hl{\textbf{RQ2}: How does disentanglement of user interests help achieve OOD generalization for CTR prediction and benefit explainable recommendation?}
    
    \item \textbf{RQ3}: What is the effect of each component in DiseCTR?
    
\end{itemize}
\subsection{Experimental Settings}
\subsubsection{Datasets}
We utilize the benchmark Amazon\footnote{\url{https://www.amazon.com/}} dataset and real-world datasets collected from two largest short-video platforms in China, WeChat Channels\footnote{\url{https://www.wechat.com/}} and Kuaishou\footnote{\url{https://www.kuaishou.com/}}.
Table \ref{tab::dataset} shows the statistics of the adopted datasets.
The details of the datasets for evaluation are as follows:
\begin{itemize}
    \item \textbf{Amazon}: This is a public benchmark dataset\footnote{\url{https://nijianmo.github.io/amazon/index.html}} for CTR prediction \citep{ni2019justifying}, which contains product reviews written by customers on the Amazon e-commerce website.
    We split the dataset into training, validation and test sets according to timestamps with the ratio of 8:1:1.
    The adopted features are: ReviewerID, ItemID, ItemCategory, ItemBrand and Price.
    
    \item \textbf{Wechat}: This is a public dataset released by WeChat Big Data Challenge 2021\footnote{\url{https://algo.weixin.qq.com/}}, which contains the logs on Wechat Channels (short-video platform) within 14 days.
    We use the data of first 10 days, middle 2 days and last 2 days as training, validation and test set.
    We use the following features: UserID, VideoID, DeviceID, AuthorID, BGMSongID, BGMSingerID, Duration, UserActivness, VideoPopularity.

    \item \textbf{Kuaishou}: This is an industrial dataset which contains the user interactions with short videos on Kuaishou APP during March 6 to March 15 in 2021.
    We use the data till March 13 as training set.
    We utilize the data on March 14 for validation, and evaluate the final performance with the data on March 15.
    We use the following features: UserID, VideoID, AuthorID, VideoCategory, AuthorCategory, Duration, UserActiveness, VideoPopularity.
\end{itemize}

\begin{table}
\centering
\caption{Statistics of the adopted datasets.}
\label{tab::dataset}

\begin{tabular}{cccc}
      \toprule
      Dataset & Users & Items & Instances \\
      \midrule
      Amazon & 84,260 & 385,897 & 3,278,991 \\
      Wechat & 20,000 & 96,488 & 7,314,206 \\
      Kuaishou & 20,303 & 222,791 & 10,848,234 \\
      \bottomrule
    \end{tabular}

\end{table}

\begin{table}
\centering
\caption{OOD evaluation on a single affected feature.}
\label{tab::ood}
\begin{tabular}{c|cc|cc}
      \toprule
      \multicolumn{1}{c|}{\multirow {2}{*}{$X_k$}} & \multicolumn{2}{c|}{Train} & \multicolumn{2}{c}{Test} \\
      \multicolumn{1}{c|}{} & Y=1 & Y=0 & Y=1 & Y=0 \\
      \midrule
      Cheap/Short & $e$ & $1 - e$ & $e'$ & $1 - e'$ \\
      Expensive/Long & $1 - e$ & $e$ & $1 - e'$ & $e'$ \\
      \bottomrule
    \end{tabular}

\end{table}

\subsubsection{OOD Evaluation}
To evaluate the generalization ability, the distributions of training and test data need to be different.
We consider two common OOD issues in real-world scenarios as follows,
\begin{itemize}
    \item \textbf{OOD-easy.} The distribution variation mainly occurs on one single feature $X_k$, such as price in e-commerce recommendation when users are given coupons.
    Thus the differences of $P(X,Y)$ between training and test data mostly come from the variation of $P(Y|X_k)$, while other distributions $P(Y|X_i)$ remain largely unchanged.
    For simplicity, we select price and video duration as the affected feature $X_k$ for different datasets, and transform the training and test data following the standard protocol introduced in related literature investigating OOD generalization \citep{arjovsky2019invariant,lu2021nonlinear}.
    Specifically, the click-through rate for cheap/short and expensive/long items are different in training and test data.
    As illustrated in Table \ref{tab::ood}, we set click-through rate $e$ as 0.6 in the training set, while we vary $e'$ to obtain multiple test sets with different intensities of distribution variation.
    As shown in Table \ref{tab::ood}, the click-through rate of the single affected feature needs to be different in training and test data.
    To achieve such goal, we perform data sampling from the original datasets.
    Specifically, suppose the CTR of the cheap products in the Amazon dataset is 0.6. To obtain an OOD dataset with 0.2 CTR of cheap products, we randomly select 1/6 of the clicked cheap samples, and combine them with all the unclicked cheap samples. Then, CTR of cheap products is (0.6*1/6)/(0.6*1/6+0.4)=0.2.
    The data we use, including user, item, features, and clicks, are all from real-world applications.
    The Amazon and Wechat datasets are widely used as open benchmarks and Kuaishou dataset is from a real-world recommender platform.
    We did not synthesize the click values, and only use different sample ratios to obtain OOD data. 
    In summary, we only perform random sampling to obtain OOD datasets, and all the clicked and unclicked samples are real.
    
    \item \textbf{OOD-hard.} We use multiple types of behaviors in the datasets, including click and like.
    Specifically, we train the model with click-behavior data, and evaluate the performance of predicting like-behavior.
    It is more challenging than OOD-easy since $P(Y|X)$ changes drastically, with $P(Y_{like}=1|X)$ much smaller than $P(Y_{click}=1|X)$.
    Note that the generalization ability for OOD-hard is critical in practice, because models are prone to overfitting when trained to predict like behavior, due to the insufficient training samples.
    Generalizing from rich click data to scarce like data can largely eliminate the overfitting issue.
\end{itemize}
In both cases, we follow the evaluation settings in \citep{bengio2019meta} to measure the generalization ability.
Specifically, after convergence on the IID training data, we transfer all the models by fine-tuning with a small fraction (10\%) of OOD data, and investigate the performance on OOD data.

\subsubsection{Baselines and Metrics}
We compare DiseCTR with state-of-the-art approaches for CTR prediction.
The details of the adopted baselines are as follows,
\begin{itemize}
    \item \textbf{FM} \citep{rendle2010factorization}: This method captures feature interaction by taking inner product of feature embeddings.
    We use the most adopted 2-order FM which learns a cross feature for each pair of input features.
    \item \textbf{DeepFM} \citep{guo2017deepfm}: This method is an ensemble model which combines FM and Deep Neural Networks (DNN) to capture both second-order and higher order feature interactions.
    \item \textbf{NFM} \citep{he2017neural}: This method proposes a Bi-Interaction layer to replace the inner product operation in FM, which increases the non-linearity modeling of high order feature interactions.
    \item \textbf{AutoInt} \citep{song2019autoint}: This is the state-of-the-art method which utilizes a multi-head self-attentive neural network to automatically learn feature interactions.
    \item \textbf{AFN} \citep{cheng2020adaptive}: This is the state-of-the-art approach which designs an adaptive factorization network to learn arbitrary-order feature interactions.
    \item \textbf{DESTINE} \citep{zhu2021disentangled}: This method proposes a self-attentive neural network to disentangle pairwise feature interaction and unary feature importance.
\end{itemize}
We do not include AutoFIS \citep{liu2020autofis} due to its high computational cost.
For fair comparison, we do not include those models that take users' historical interaction sequence as input like DIN \citep{zhou2018deep} and DIEN \citep{zhou2019deep}.
It is worth noting that DiseCTR is fully compatible with sequential models and we leave it for future work.
The two OOD recommendation methods \cite{he2022causpref, wang2022causal} are not included since they can not handle the general OOD-easy and OOD-hard problems in CTR prediction.
To measure the prediction accuracy, we utilize two widely adopted metrics, AUC and GAUC \citep{zhou2018deep,chang2021sequential}.

\subsubsection{Experiment Setups}
For both OOD-easy and OOD-hard experiments, we first train the recommendation model with IID data.
We use Adam \citep{kingma2014adam} to update the model parameters, and the learning rate is set as 0.0001.
The adopted loss function is a combination of BPR loss \cite{rendle2012bpr} and the proposed discrepancy loss as shown in Eqn (\ref{eq::loss}), and we set the loss weight $\alpha$ as 0.1.
We use a single Nvidia GeForce GPU for model training, and the batch size is 2048.
We train the models with early-stopping to avoid over-fitting.
Specifically, we stop training when the AUC on validation set does not make progress for 5 epochs.
After convergence on the IID training data, we finetune each model with a small fraction (10\%) of OOD data.
Finally, we evaluate the recommendation performance on OOD test data to compare the generalization ability of different models.
To guarantee fair comparison, we fix the embedding size as 32 for all methods, and other hyper-parameters are carefully tuned by grid search.
For DiseCTR, we set the number of interests $M$ as 4, and the number of attention heads $H$ and $h$ in the encoder and aggregator as 2 and 8, respectively.
Code based on TensorFlow \citep{abadi2016tensorflow} and datasets to reproduce the experimental results in this paper are released at \url{https://github.com/DavyMorgan/DiseCTR/}.

\subsubsection{Weak supervision for OOD-easy}
For OOD-easy, pairwise weak supervision is enhanced to make the last interest capture the affected feature.
Specifically in OOD-easy, distribution shift occurs on feature $X_k$ which makes $P(Y|X_k)$ different in training and test data.
Since the conditional distributions of other features are largely invariant, we propose to separate the interest related to $X_k$ from other interests.
Specifically, we force the last (\textit{i.e. $M$-th}) embedding $\mathbf{z}_M$ in $\mathbf{Z}$ to capture the interest about $X_k$, and $\mathcal{A}$ is decided according to $X_k$ in the pair:
\begin{equation}
    (\mathcal{A}, \bar{\mathcal{A}}) = 
    \begin{cases} 
        (\{M\}, \{1, 2, \cdots, M-1\}), & \mbox{if } X^{pos}_k = X^{neg}_k \\
        (\varnothing, \{1, 2, \cdots, M\}),  & \mbox{if } X^{pos}_k \ne X^{neg}_k
    \end{cases}
    \label{eq::easy_weak}
\end{equation}
In other words, we take average of the $M$-th interest if $X_k$ in the positive sample is the same with the negative sample.

To further regularize the meanings of $\mathbf{Z}$, we utilize a fully connected classifier $\mathbf{W_{weak}} \in \mathbb{Z}^{N_k \times d'}$ to reconstruct $X_k$ from $\mathbf{Z}$, inspired by recent advances in disentangled VAE \citep{locatello2019disentangling}.
Specifically, we expect $X_K$ to be predictable from $\mathbf{z}_M$, while not predictable from other interests.
Therefore, we apply the extra prediction task on $\mathbf{z}_M$, as well as $\mathbf{z}_i$ with $i < M$ but in an adversarial way :
\begin{align}
    &\hat{X}_k^{(i)} = \mathrm{SoftMax}(\mathbf{W_{weak}} \cdot \mathbf{z}_i),\\ 
    &L_{weak} = L_{CE}(X_k, \hat{X}_k^{(M)}) - \sum_{i=1}^{M-1}{L_{CE}(X_k, \hat{X}_k^{(i)})},
\end{align}
where $ L_{CE}$ is the CrossEntropy loss function, and the adversarial negative loss on $\mathbf{z}_i$ with $i < M$ is implemented with a gradient reversal layer \citep{ganin2015unsupervised,zheng2021dgcn}.
By mapping the the $M$-th interest to the affected feature $X_k$, we largely reduce the impact of distribution shift, since only one sub-encoder is influenced while other sub-encoders are to great extent stable in OOD cases.

It is worthwhile to notice that baseline methods learn rather entangled interest representations, which means that the influence of the affected feature $X_k$ is almost contained in every learned interest embedding.
Thus the impact of $X_k$ can not be separated to one specific interest in the baselines.
Intuitive solutions like discarding the affected features $X_k$ is investigated in Section \ref{exp::calibration}, however, the performance is even worse since useful information of $X_k$ is also removed.

In real-world scenarios, OOD-easy cases are quite common.
For example, users can grow loyalty on a specific brand, then the brand-aspect interest varies while other interests may largely remain stable.
Another example can be providing coupons to users which almost only changes the economic-aspect interest, as introduced in the paper.
Being capable of adapting to specific OOD issues is a great advantage of the proposed DiseCTR, and such advantage comes from the disentanglement design and the effective usage of the partial-variation property.
Existing baselines can not handle OOD-easy cases well since different interests are entangled with each other and the impact of the affected feature $X_k$ can not be captured separately.

In fact, learning disentangled interests makes it possible for DiseCTR to separate the influence of distribution variation in OOD-easy.
Baseline methods are limited in eliminating the impact of distribution shift.
One potential solution is data calibration such as removing the affected $X_k$ from input features.
However, the performance is even worse and we show the results in Section \ref{exp::calibration}.

\subsection{Generalization Performance (RQ1)}

\begin{table*}[t]
    \caption{Overall performance comparison. We evaluate the accuracy after finetuning with 10\% of OOD data. Two evaluation protocols, OOD-easy and OOD-hard are adopted. \underline{Underline} means the 2nd and 3rd best methods, \textbf{bold} means \textit{p}-value < 0.05.}
    \label{tab::overall}
    \begin{tabular}{c|cc|cc|cc|cc|cc}
      \toprule
      Protocol & \multicolumn{6}{c|}{OOD-easy ($e=0.6, e'=0.2$)} & \multicolumn{4}{c}{OOD-hard (click $\rightarrow$ like)}\\
      \hline
      Dataset & \multicolumn{2}{c|}{Amazon} & \multicolumn{2}{c|}{Wechat} & \multicolumn{2}{c|}{Kuaishou} & \multicolumn{2}{c|}{Wechat} & \multicolumn{2}{c}{Kuaishou}\\
      \hline
      Method & AUC & GAUC & AUC & GAUC & AUC & GAUC & AUC & GAUC & AUC & GAUC  \\
      \midrule
    FM  & 0.5915 & 0.4694 & 0.8356 & 0.7495 & 0.6901 & 0.5034 & \underline{0.6880} & 0.5396 & 0.7256 & 0.5037 \\
      DeepFM  & 0.6273 & 0.5156 & 0.7954 & 0.7232 & 0.7064 & 0.5452 & 0.6710 & \underline{0.5400} & \underline{0.7547} & \underline{0.5356} \\
      NFM  & \underline{0.6736} & \underline{0.5815} & 0.8336 & 0.7433 & 0.6955 & 0.5278 &  0.6716 & 0.5237 & 0.7424 & 0.5154 \\
      AutoInt  & 0.6279 & 0.5108 & \underline{0.8362} & 0.7508 & \underline{0.7113} & \underline{0.5500}  & 0.6686 & 0.5387 & 0.7252 & 0.5051\\
      AFN  & \underline{0.6614} & \underline{0.5676} & \underline{0.8387} & \underline{0.7557} & 0.7086 & 0.5425 & \underline{0.6850} & \underline{0.5442} & \underline{0.7636} & \underline{0.5251} \\
      DESTINE  & 0.6344 & 0.5277 & 0.8335 & \underline{0.7520} & \underline{0.7100} & \underline{0.5482} & 0.6711 & 0.5356 & 0.6792 & 0.4820 \\
      DiseCTR & \bf{0.6762} & \bf{0.5972} & \bf{0.8532} & \bf{0.7678} & \bf{0.7198} & \bf{0.5736} & \bf{0.7071} & \bf{0.5494} & \bf{0.7653} & \bf{0.5449} \\
      \bottomrule
    \end{tabular}
\end{table*}

\subsubsection{\textbf{Overall Comparison.}}
\begin{figure}[t]
    \centering
    \includegraphics[width=\linewidth]{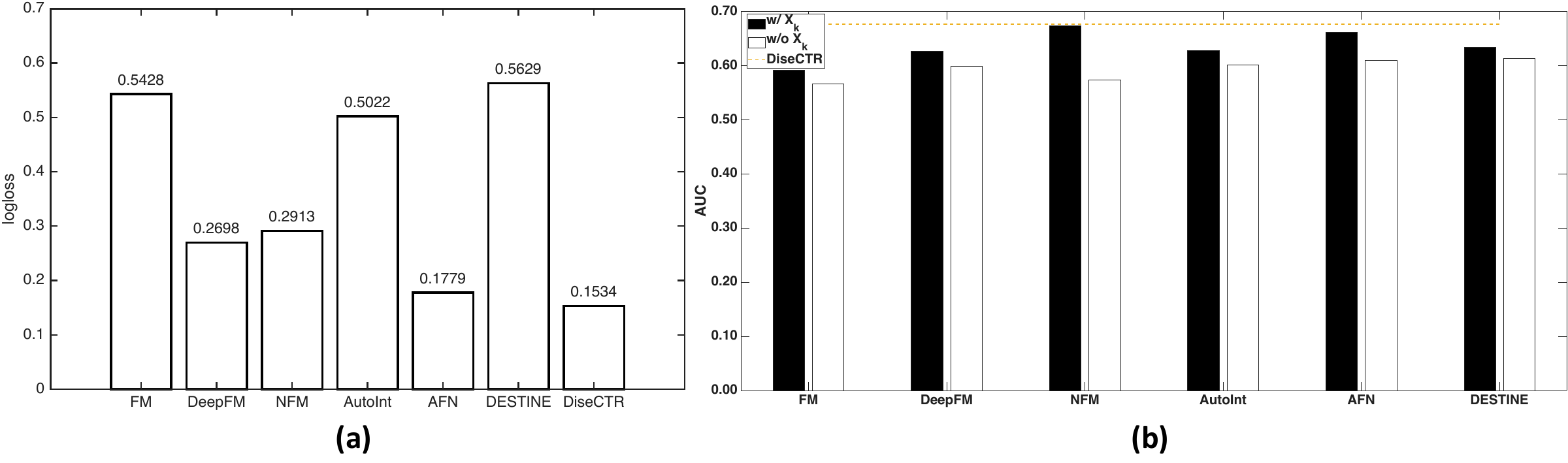}
    \caption{(a) Logloss of different methods on the Kuaishou dataset under OOD-hard protocol. (b) AUC of baseline methods with the affected feature (price) discarded on Amazon dataset.}
    \label{fig::discarding}
\end{figure}

We set $e'$ as 0.2 that is different from $e=0.6$ in training data to evaluate OOD-easy performance, and use \textit{like} behavior as the OOD environment against \textit{click} behavior for OOD-hard.
Table \ref{tab::overall} shows the performance on three datasets.
Note that the Amazon dataset only has one type of behavior, and thus we only evaluate OOD-easy performance on the Amazon dataset.
From the results we have the following observations:
\begin{itemize}
    \item \textbf{Directly learning correlations of low-level features performs poorly in OOD cases.}
    FM, DeepFM and NFM, provide much worse performance in most cases compared with other methods.
    These methods concatenate all the input features, and utilize rather simple feature interaction layers, like inner product and multi-layer perceptrons, for prediction.
    Since directly learning $P(Y|X)$ is vulnerable to the distribution variation, these methods perform poorly in OOD scenarios.
    \item \textbf{Entangled modeling of interests is insufficient to accomplish OOD generalization.}
    Although AutoInt and AFN leverage self-attention and adaptive feature interaction, which are large-capacity neural layers, compared with direct matching raw features, they still cannot capture the multiple aspects of interests.
    In other words, the obtained embeddings are redundant with much mutual information, which entangle different aspects.
    Therefore, their performance in OOD cases is still limited since distribution variation can influence a large part of these models.
    \item \textbf{DiseCTR can generalize well to OOD scenarios.}
    Our DiseCTR achieves the best OOD performance in most cases with significant improvements.
    Specifically, on Amazon dataset, GAUC is improved by 0.016 against NFM; DiseCTR improves AUC by about 0.015 (OOD-easy) and 0.020 (OOD-hard) against AFN on Wechat dataset; on Kuaishou dataset, GAUC is improved by 0.024 (OOD-easy) and 0.040 (OOD-hard) compared with AutoInt.
    \hl{We also calculate the logloss~\cite{yu2023cognitive} metric of all methods for the challenging OOD-hard scenario on Kuaishou dataset.
    As illustrated in Figure \ref{fig::discarding}a, our DiseCTR approach achieves the lowest logloss with over 13.7\% improvements against AFN.}
    By learning disentangled representations of interests, DiseCTR reduces the impact of distribution shift to only a small fraction of the model, achieving fast transfer towards unseen distributions.
\end{itemize}

\subsubsection{\textbf{Performance of Data Calibration}}\label{exp::calibration}
One intuitive solution to OOD-easy is simply discarding the affected feature $X_k$, then the distribution variation of remaining features will be largely eliminated.
However, such straightforward approach also removes useful information which may lead to worse performance.
For example, providing users with coupons change the distribution $P(Y|X_{price})$, while removing $X_{price}$ from the input features will make the model fail to learn the drifted relation from new data, thus it can not recommend proper items with higher price.
We discard the affected feature for all the baselines, and results are shown in Figure  \ref{fig::discarding} (b).
We can observe that the performance of all the baselines drops drastically after removing the price feature of Amazon dataset.
In other words, it is better to \textit{adapt} to the varied distribution $P(Y|X_{k})$ instead of \textit{ignoring} it.
Through interest disentanglement, the proposed DiseCTR can achieve fast adaptation to new distributions.

\begin{figure}[t]
    \centering
    \includegraphics[width=\linewidth]{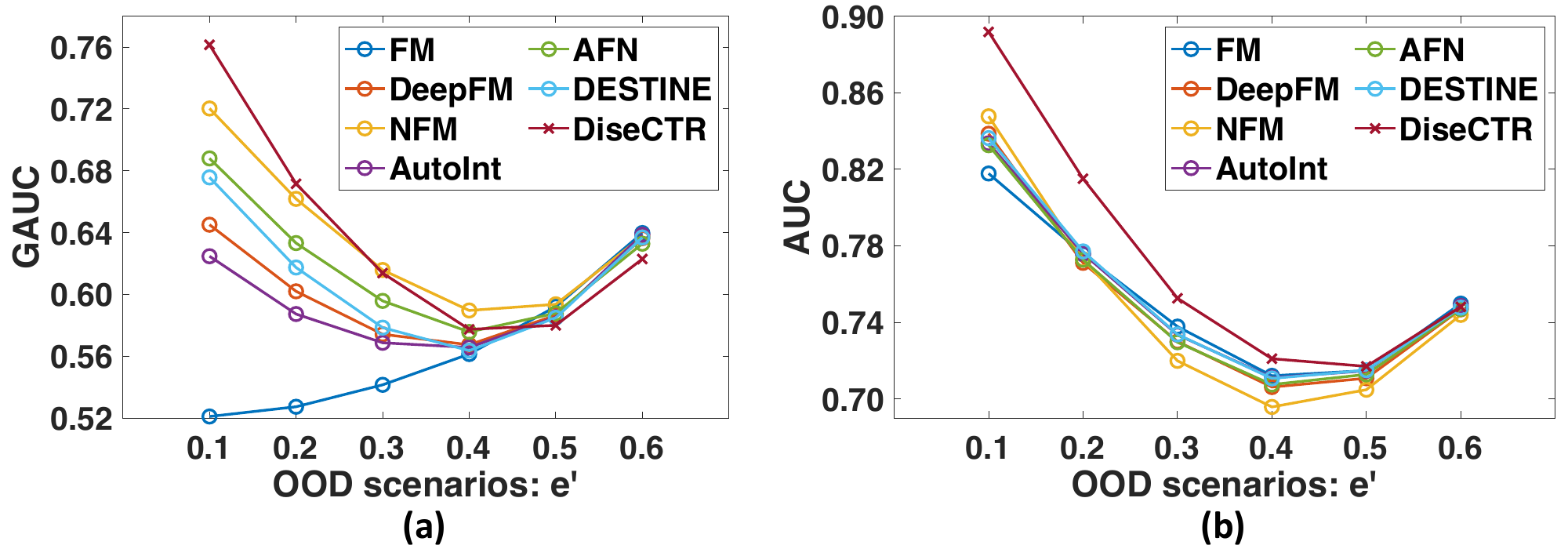}
    \caption{Transfer accuracy evaluation on (a) Amazon and (b) Kuaishou dataset. Higher means better.}
    \label{fig::transfer_ood}
\end{figure}

\begin{figure}[t]
    \centering
    \includegraphics[width=\linewidth]{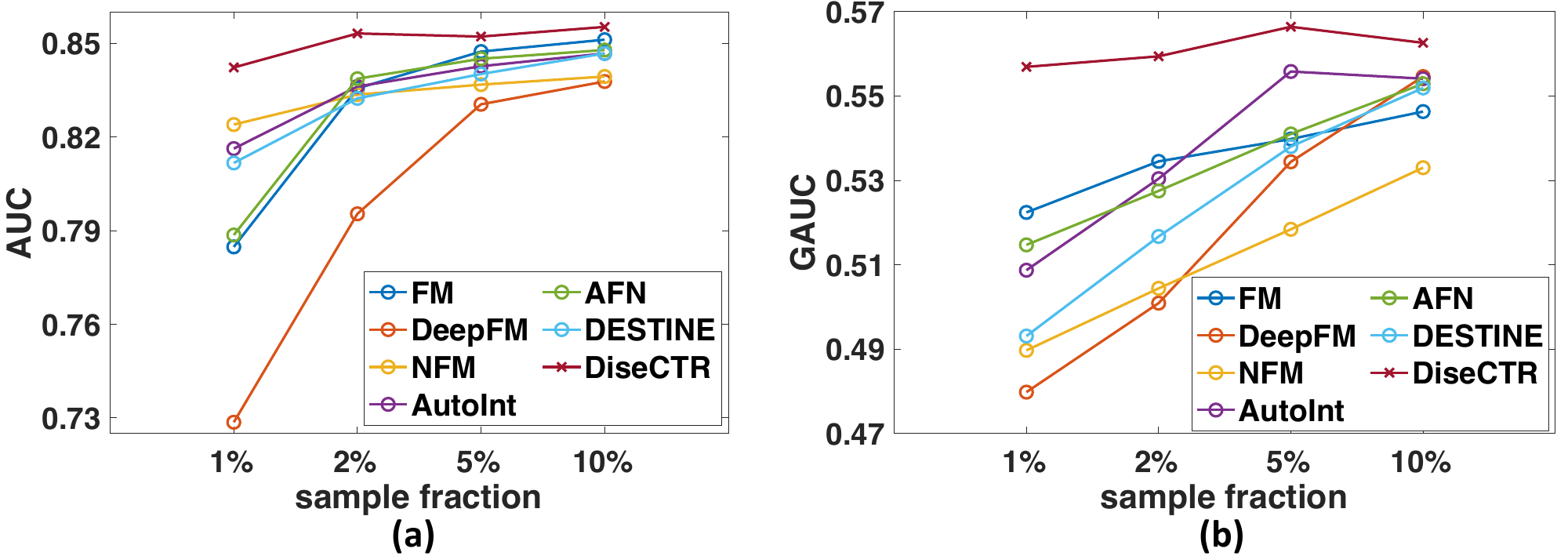}
    \caption{Transfer efficiency comparison. (a) OOD-easy on Wechat dataset. (b) OOD-hard on the Kuaishou dataset.}
    \label{fig::transfer_efficiency}
\end{figure}

\subsubsection{\textbf{Transfer Accuracy on Different OOD Scenarios.}}
Since the intensity of OOD issue is usually unknown in practice, we vary the value of $e'$ in the range of 0.1 and 0.6 to simulate multiple environments with different intensities of OOD issue.
Intuitively, smaller $e'$ indicates larger distribution variation with training data.
Meanwhile, as $e'$ decreases, the affected feature $X_k$ also becomes a stronger predictor in the OOD scenario.
Figure \ref{fig::transfer_ood} illustrates the results on Amazon and Kuaishou dataset.
We can find that all the models are of similar performance when $e'=0.6$ which equals to $e$ in training data, where the IID assumption holds true.
However, more importantly, as we decrease $e'$ to simulate OOD scenarios, the proposed DiseCTR starts to outperform other baselines.
Specifically, we can observe that DiseCTR outperforms other baselines in almost all levels of distribution variation.
Meanwhile, the improvements become larger as $e'$ moves to extreme OOD cases, where AUC on the Kuaishou dataset is improved by over 0.02 when $e'=0.3$, and over 0.04 when $e'\leq0.2$.
By learning disentangled representations of multiple interests, DiseCTR is more stable in OOD scenarios since distribution variation only influences a few interests and the majority of the model can still take effect.
In onther words, only a few parameters related to the affected feature need to be updated significantly, while other parameters only need a small amount of fine-tuning.
Therefore, DiseCTR can effectively adapt to multiple OOD scenarios with different intensities of distribution variation.

\subsubsection{\textbf{Comparison of Transfer Efficiency.}}
Since OOD data is usually expensive and hard to obtain in practice, models are required to transfer with limited data.
Therefore, we use different amounts of OOD data for fine-tuning to investigate the transfer efficiency, which has also been adopted in \citep{bengio2019meta}.
Figure \ref{fig::transfer_efficiency} demonstrates the performance of fine-tuning with 1\%, 2\%, 5\%, and 10\% of training data.
We can find that DiseCTR consistently outperforms baselines under both protocols, even with only 1\% of the OOD training data, and the improvements are consistently significant with different amounts of OOD training data.
Meanwhile, the performance with 1\% sample fraction is very close to the performance with 10\% sample fraction, which means that DiseCTR only requires a very small number of samples to accomplish transfer towards unseen distributions.
Overall speaking, the disentangled structures of interest largely reduce the amount of parameters that need to be updated, which makes DiseCTR achieve fast transfer with high data efficiency.

\hl{\subsection{Explainability and Disentanglement Analysis (RQ2)}}

\begin{figure}[t]
    \centering
    \includegraphics[width=\linewidth]{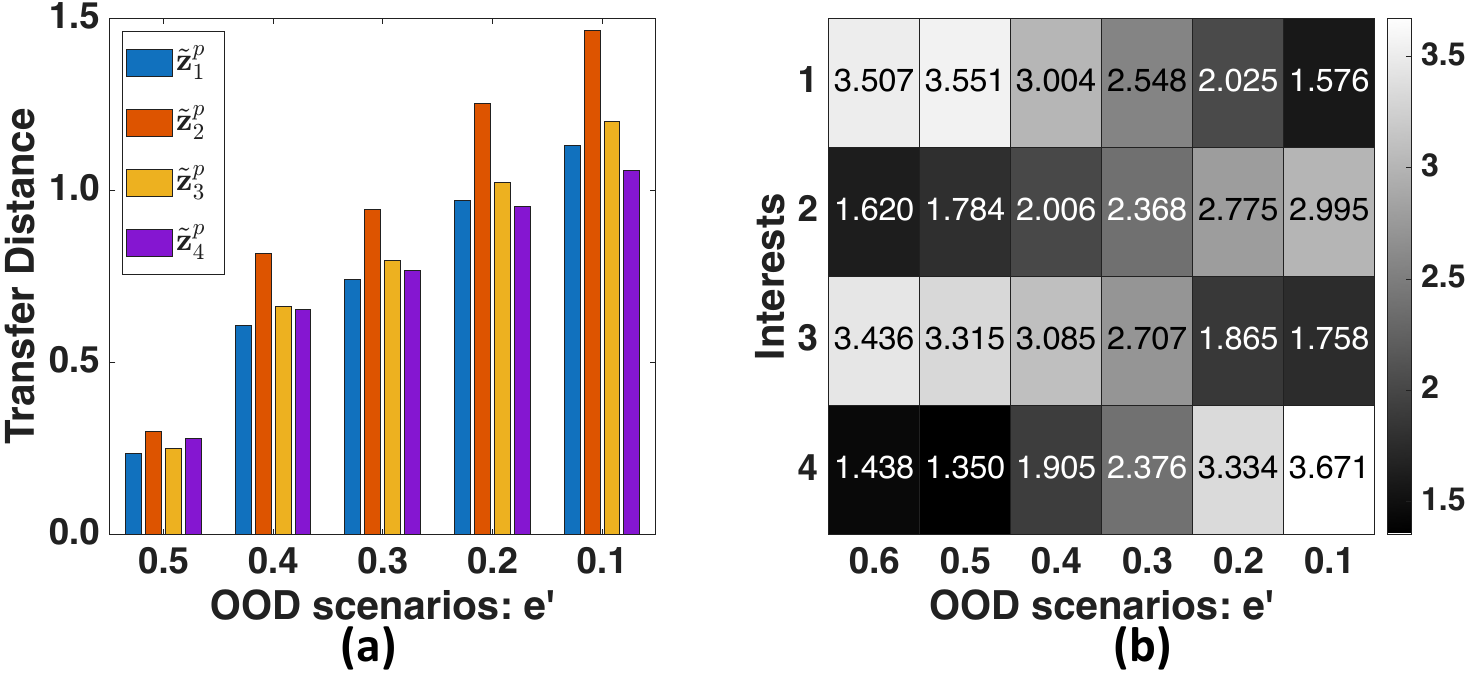}
    \caption{(a) Transfer distance of prototypes on Kuaishou dataset. (b) Heatmap of attention weights on different interests on Kuaishou dataset.}
    \label{fig::disen}
\end{figure}

\hl{We analyze the generalization ability of DiseCTR and how it benefits explainable recommendation from the perspective of model parameters.
Specifically, we visualize the learned interest embeddings to demonstrate the independence of different interests, which is a critical aspect for explainability.}
To empirically verify that leveraging the partial-variation property is critical in OOD generalization, we investigate how the parameters in DiseCTR change during the transfer process.
Specifically, we calculate the \textit{transfer distance} of interest prototypes, which is the value differences between parameters before and after fine-tuning with OOD data.
Note that larger distance indicates more significant updates.
Figure \ref{fig::disen}(a) illustrates the transfer distances to different OOD scenarios with various $e'$.
We can observe that the transfer distances for all the four prototypes increase as we gradually decrease $e'$, which is in line with expectations since smaller $e'$ indicates larger distribution shift.
Particularly, the transfer distance of one prototype $\tilde{\mathbf{z}}_2^p$ is much larger than the other three prototypes.
In other words, only a small fraction of parameters in DiseCTR need significant updating, while most of the parameters only need small modification.
\hl{Meanwhile, the results in Figure \ref{fig::disen}(a) also indicates that the interest shift and evolution are mostly captured by the prototype embedding $\tilde{\mathbf{z}}_2^p$, and other prototype embeddings $\{\tilde{\mathbf{z}}_1^p, \tilde{\mathbf{z}}_3^p, \tilde{\mathbf{z}}_4^p\}$ capture the unchanged user interests, which significantly improve the explainability of recommendation.
By learning disentangled interest embeddings, DiseCTR leverages the partial-variation property, which accomplishes explainable and fast transfer to OOD cases.}

We also visualize the attention scores in the interest aggregator before and after fine-tuning with OOD data.
We compute the rank of four interests according to their attention scores, where larger rank of one interest indicates that it gains more attention against other interests.
Specifically, we take one attention head and calculate the attention score for 4,096 samples.
We then rank the attention scores of four interests and calculate the averaged ranks of these 4,096 samples.
Notice that we force the last interest embedding to capture the interest related to the affected feature (video duration).
Meanwhile, $e'$ indicates the positive sample rate of short videos, thus $e'=0.5$ is the most difficult case to predict CTR from video duration, while $e'=0.1$ is the easiest case where video duration is a very strong feature.
Figure \ref{fig::disen}(b) demonstrates how attention changes from the IID case ($e'=0.6$) to different OOD cases on the Kuaishou dataset.
We can observe that the attention score of the last interest drops when $e'=0.5$, and then continuously increases as $e'$ turns down towards extreme OOD cases.
In other words, the attention score on the last interest greatly matches the feature importance of video duration, which verifies that DiseCTR successfully disentangles different interests.
\hl{More importantly, the consistent trends of test data and attention rank observed in Figure \ref{fig::disen}(b) imply that the 4-th interest captures the user preferences on video duration, and the first 3 interests capture the user preferences on other factors unrelated to video duration, thus facilitating explainable recommendation using the attention ranks of different interests.}

\subsection{Ablation and Hyper-parameter Study (RQ3)}

\begin{figure}[t]
    \centering
    \includegraphics[width=0.85\linewidth]{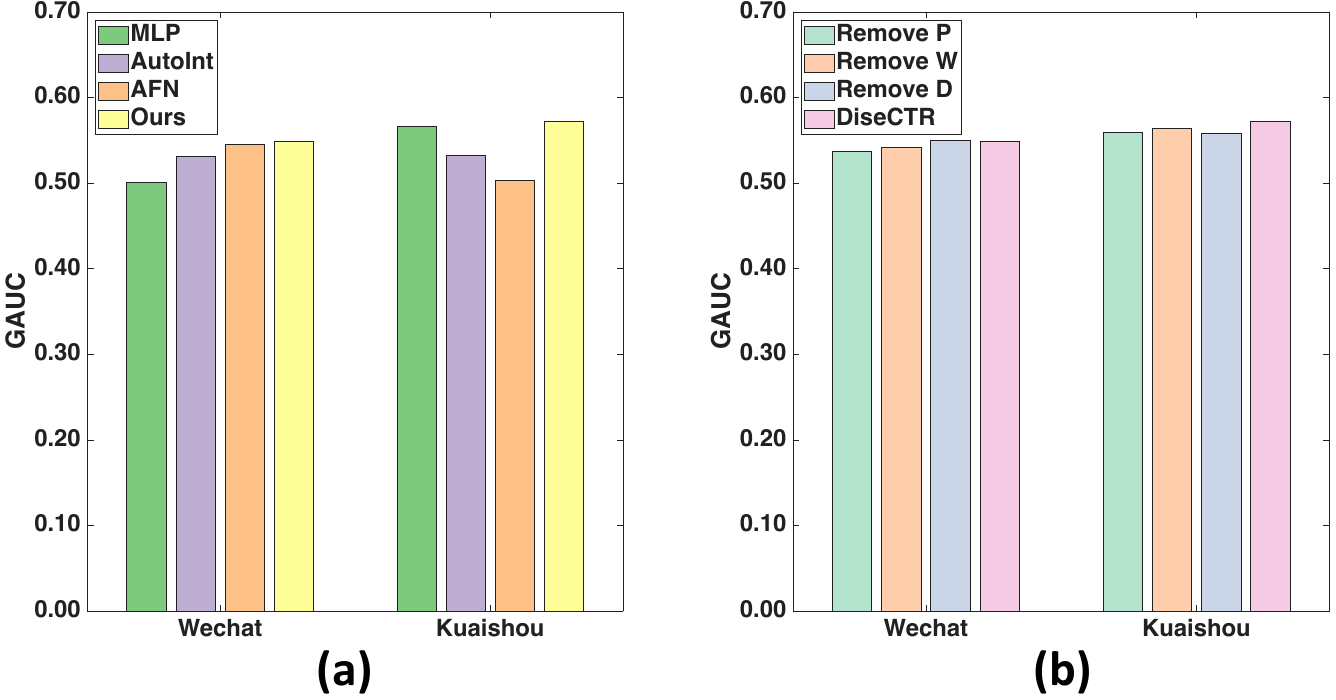}
    \caption{(a) Performance comparison of different encoders. (b) Ablation on interest disentangler.}
    \label{fig::ablation}
\end{figure}

\begin{table}[t]
    \caption{Results of different $M$ on Wechat dataset.}
    \label{tab::hyper}
    \centering
    \begin{tabular}{c|cccc}
      \toprule
      $M$ & 1 & 2 & 4 & 8 \\
      \midrule
      AUC & 0.6911 & 0.6931 & \bf{0.7071} & 0.6871 \\
      \bottomrule
    \end{tabular}
\end{table}

We verify the design of DiseCTR by comparing with several counterparts of interest encoder and removing specific components of interest disentangler.
\subsubsection{\textbf{Effectiveness of Interest Encoder.}}
In the encoder, sparse attention forces each interest to only focus on a few input features.
We compare with other encoders including MLP, AutoInt \citep{song2019autoint} and AFN \citep{cheng2020adaptive}.
Figure \ref{fig::ablation}(a) illustrates the results on two datasets.
Dense connections like AutoInt provide worse performance since each interest is related to almost all features, which is vulnerable to distribution variation.
Our encoder achieves the best GAUC on both datasets, which confirms the necessity of constructing sparse connections between interests and features.

\subsubsection{\textbf{Effectiveness of Interest Disentangler.}}
We investigate different components of the disentangler.
Figure \ref{fig::ablation}(b) shows the performance of DiseCTR and its variants with specific designs removed including prototypes (P), weak supervision (W) and discrepancy loss (D).
We can observe that removing prototypes leads to the worst GAUC on the WeChat dataset and the second worst GAUC on the Kuaishou dataset.
Without prototype vectors, interest clustering is conducted directly on the encoder outputs, which have high variance.
The prototypes provide learnable cluster centroids, which reduces the variance of interest embeddings greatly after clustering.
Moreover, pairwise weak supervision averages interests of sample pairs adaptively, which forces interest embeddings to capture shared and distinguishing interests respectively.
We can find that removing it leads to worse performance on both datasets.
At last, the discrepancy regularization encourages the independence between different interests, and removing $L_{dis}$ results in drastic performance drop on Kuaishou dataset.

\subsubsection{\textbf{Impact of the Number of Interests.}}\label{app::hyper}
We compare the performance of DiseCTR with different number of interests.
Table \ref{tab::hyper} shows the results of OOD-hard on Wechat dataset.
We can observe that increasing $M$ from 1 to 4 can effectively improve the expressive ability of DiseCTR and achieve better performance.
In addition, further increasing $M$ to 8 may introduce redundancy and overfitting, which leads to performance drop.

\vspace{10px}
In summary, we conduct extensive experiments to demonstrate the OOD generalization ability of the proposed method.
On the one hand, we show that DiseCTR can generalize well to different OOD scenarios, and the improvements are particularly larger in extreme OOD cases.
On the other hand, we illustrate that DiseCTR can achieve OOD generalization with high data efficiency.
Furthermore, we show that leveraging the partial-variation property is the key of OOD generalization. 
Specifically, by disentangling user interests, only a small number of parameters related to the varying interest need a significant update, while most of the parameters only need a small amount of fine-tuning, which makes DiseCTR generalize towards unseen distributions accurately and efficiently.

\section{Related Work}\label{sec::related}

\subsection{Interest Learning in CTR Prediction}
Learning user interests is critical for CTR prediction, while most studies learn a relatively entangled interest representation.
Specifically, existing methods \citep{rendle2010factorization,he2017neural,guo2017deepfm,xiao2017attentional,lian2018xdeepfm,cheng2016wide,xu2020learning,shan2016deep,liu2020autofis,qin2020user,pi2019practice,zhou2018deep,zhou2019deep,zhu2021disentangled,yu2020deep,zhang2016collaborative,wang2015collaborative,wang2016collaborative,qin2023learning,10.1145/3626772.3657777,yu2023cognitive} aim to perform direct matching from features to interactions.
For example, Rendle \textit{et al.} \cite{rendle2010factorization} utilized inner product to capture feature interaction between every feature pair.
The inner product operation is further replaced by a Bi-interaction layer in \cite{he2017neural}, and combined with MLP in \cite{guo2017deepfm}.
Recently, Song \textit{et al.} \cite{song2019autoint} proposed to learn interests based on self-attention, and Cheng \textit{et al.} \cite{cheng2020adaptive} further developed an adaptive factorization network to learn arbitrary-order feature interactions.
\hl{Besides pre-defined feature interactions, \citet{yu2023cognitive} proposed an evolutionary search approach to adaptively select proper interactions on features for accurate CTR prediction.} 
However, they largely ignore $Z$ and mainly capture $P(Y|X)$, which can be vulnerable to distribution shift.
Unlike existing methods that generalize from one data point to another one in the same distribution, we focus on the generalization from one distribution to unseen distributions and propose to learn disentangled interests to achieve such a goal.

\hl{\subsection{Disentangled Recommendation}
Some studies \citep{zheng2021disentangling,zhang2021causal,chen2021autodebias,wang2020disentangled,wang2021learning,zhu2021disentangled,wang2023intent,zhang2024denoising,ren2023disentangled,zheng2022disentangling} investigate disentangled user interests in recommendation systems.
For example, \citet{wang2023intent} proposed to learn independent representations for different user intents by contrastive learning on a user-item-concept graph.
\citet{zhang2024denoising} divided an item graph into multiple subgraphs with distinct semantics, then leveraged graphical disentangled learning to capture user's diverse preferences within each subgraph.
\citet{ren2023disentangled} further improved the robustness of user interest disentanglement and distilled fine-grained latent factors with adaptive self-supervised augmentation.
However, most of these works only tackle the simplified collaborative filtering (CF) task, and they are all based on the IID assumption which hardly holds true in real-world scenarios.
Two recent works \citep{wang2022causal,he2022causpref} investigate the OOD issue in recommendation.
Nevertheless, \citet{he2022causpref} studies OOD in CF and can not handle the more complicated CTR prediction task.
Meanwhile, \citet{wang2022causal} only investigates user feature shift such as income increase, which can not solve the more general OOD-easy and OOD-hard tasks in this paper.
Different from these methods, our work investigates general OOD scenarios in the more complicated CTR prediction task, which is quite common in real applications.}

\subsection{Model Robustness and OOD Generalization}
Machine learning theories guarantee optimum accuracy in IID cases, while the performance often drops drastically in OOD scenarios \citep{mohri2018foundations,scholkopf2021toward}.
To obtain a robust model that generalizes to unseen distributions, several studies \citep{arjovsky2019invariant,lu2021nonlinear,xie2020n,zhang2021deep,shen2020stable,lu2022invariant,shen2021towards} were proposed to remove spurious correlations in the training data.
For example, \citet{arjovsky2019invariant} proposed invariant risk minimization (IRM) which can guarantee generalization ability for linear models.
\citet{lu2021nonlinear} further extends IRM to non-linear cases with a causal inference approach.
In terms of recommendation, robustness from the perspective of OOD generalization is largely unexplored, while a few works study the robustness under adversarial attack \citep{punjabi2018robust,liu2020certifiable,wu2021fight}, which is a quite different definition of ``robust" compared with our work.
As far as we know, we are the first to investigate OOD generalization for CTR prediction, which is a critical problem since distribution variation is common in practice.

\section{Conclusion and Future Work}\label{sec::conclusion}
In this paper, we investigate the problem of OOD generalization for CTR prediction, an emerging issue in many practical online services such as e-commerce and micro-video platforms.
A novel model named DiseCTR is introduced to reduce the impact of distribution variation by disentangling the multiple aspects of user interests.
The key of DiseCTR's architecture is to impose weak supervision on disentanglement by clustering towards learnable interest prototypes and adaptively averaging the shared interest embeddings of the sample pair.
Experiments on real-world datasets demonstrate that DiseCTR can generalize well to unseen distributions.
Further analysis on DiseCTR provide valuable insights, that stronger disentanglement can achieve lower transfer distance and more interpretable attention scores.

\hl{While DiseCTR has demonstrated promising results in handling OOD issues in CTR prediction, it has some limitations that need to be considered for practical application.
First, although DiseCTR only introduced a relatively small number of additional parameters, its computational cost may increase with very large datasets. 
Specifically, as the number of features and the embedding size increase, the computational overhead of the encoder and disentangler could become significant. 
The pairwise weak supervision also adds complexity by requiring operations on pairs of samples. 
Second, the current implementation of DiseCTR is primarily focused on offline training and evaluation. 
Adapting the model to handle real-time data streams in recommendation systems presents a number of challenges, as models for real-time recommendation need to react and update their recommendations fast. 
Thus DiseCTR needs to be further optimized to quickly adapt to new data while maintaining its OOD generalization capabilities. 
Online learning strategies, such as incremental learning, can be adopted for effective real-time adaptation.}

As for future work, we plan to further incorporate sequential modeling such as MIMN \citep{pi2019practice} and UBR4CTR \citep{qin2020user} into our proposed DiseCTR.
\hl{Besides, as a general approach for OOD generalization, it is a promising direction advanced CTR prediction models that incorporate feature interaction search such as CELS~\cite{yu2023cognitive}, and we leave it as important future work.}
We also plan to evaluate DiseCTR via online A/B tests.
In real-world applications, there are numerous OOD scenarios and the multiple types of behaviors in the same domain, \textit{i.e.} the OOD-hard problem in this paper, is only one scenario.
Thus it is a promising direction for future work to investigate other OOD scenarios such as cross domain generalization.
Overall, we believe DiseCTR makes one step closer to the real-world scenarios where distribution variation is happening all the time.

\bibliographystyle{ACM-Reference-Format}
\bibliography{bibliography}

\end{document}